\documentclass[aip,jap,groupedaddress,reprint]{revtex4-1}
\usepackage{graphicx}
\usepackage{ulem}
\usepackage{amssymb}
\begin{document}


\title{Variation of field enhancement factor near the emitter tip}
 

\author{Debabrata Biswas}
\author{Gaurav Singh} 
\author{Shreya G. Sarkar}
\author{Raghwendra Kumar}

\affiliation{
Bhabha Atomic Research Centre,
Mumbai 400 085, INDIA}


\begin{abstract}
  The field enhancement factor at the emitter tip and its variation in a close neighbourhood determines
  the emitter current in a Fowler-Nordheim like formulation. For an axially symmetric emitter with a
  smooth tip, it is shown that the variation can be accounted by a $\cos{\tilde{\theta}}$ factor in appropriately
  defined normalized co-ordinates. This is shown analytically for a hemiellipsoidal emitter and
  confirmed numerically for other emitter shapes with locally quadratic tips. 
\end{abstract}






\maketitle

\newcommand{\be}{\begin{equation}}
\newcommand{\ee}{\end{equation}}
\newcommand{\bea}{\begin{eqnarray}}
\newcommand{\eea}{\end{eqnarray}}
\newcommand{\Tbar}{{\bar{T}}}
\newcommand{\En}{{\cal E}}
\newcommand{\K}{{\cal K}}
\newcommand{\GC}{{\cal \tt G}}
\newcommand{\Lop}{{\cal L}}
\newcommand{\DB}[1]{\marginpar{\footnotesize DB: #1}}
\newcommand{\q}{\vec{q}}
\newcommand{\kt}{\tilde{k}}
\newcommand{\Lopn}{\tilde{\Lop}}
\newcommand{\noi}{\noindent}
\newcommand{\ovn}{\bar{n}}
\newcommand{\ovx}{\bar{x}}
\newcommand{\ovE}{\bar{E}}
\newcommand{\ovV}{\bar{V}}
\newcommand{\ovU}{\bar{U}}
\newcommand{\ovJ}{\bar{J}}
\newcommand{\calE}{{\cal E}}
\newcommand{\ovphi}{\bar{\phi}}
\newcommand{\zt}{\tilde{z}}
\newcommand{\nuv}{\rm v}



\section{Introduction}
\label{sec:intro}

The field of vacuum nanoelectronics involves field electron emitters with sharp tips having
radius of curvature in the nanometer regime. Due to the high aspect ratio, such emitters
can have a large  field  enhancement factor, $\gamma_a$, at the apex (tip). Several models have been studied
to gain insight into the dependence of height ($h$) and apex radius
of curvature ($R_a$) on $\gamma_a$ \cite{EV2002,forbes2003,podenok,read,roveri,bonard,smith}.
Of these, the hemiellipsoid and hyperboloid emitters are analytically tractable \cite{kos,pogorelov} while the
floating sphere at plane
potential has been studied extensively but its predictions ($\gamma_a \simeq h/R_a$)
far exceed the known results for $\gamma_a$
especially for sharp emitters \cite{ZPCL,forbes2016}.
A much studied numerical model is a cylindrical post with a
hemispherical top \cite{choice} for which various fitting formulas for $\gamma_a$ exist.
A straightforward
estimate \cite{forbes2003} is $\gamma_a \simeq 0.7 (h/R_a)$  while more
eleborate ones \cite{EV2002,forbes2003,podenok,read,bonard}
are expressed as $\gamma_a \simeq a(b + h/R_a)^\sigma$ with $0.9 < \sigma \leq 1$.
The $h/R_a$ dependence of $\gamma_a$ can be expected for various other vertically
placed emitter shapes, though there are very few concrete results.

While there is some understanding of the local field enhancement at the emitter apex, its variation
in the neighbourhood of the tip is not as clear. For the hemisphere on a plane,
$\gamma(\theta) = \gamma_a \cos\theta$, where $\gamma_a = 3$ and the origin is the center of
the (hemi)sphere. For the hemiellipsoid or the hyperboloid, the local field at the emitter surface
is known, though a geometric formula analogous to the hemisphere (the $\cos\theta$ dependence) is not
known to exist. A recent numerical study \cite{bertan} on the hemiellipsoid using the Ansys-Maxwell software
includes the variation of $\gamma$ with angle $\theta$ from the center of the ellipsoid.
For a hemisphere on a cylindrical post with the origin at the center of the
hemisphere, the variation with $\theta$ was reported to be quadratic \cite{podenok} while
another study \cite{read}  found a $\cos^{1/2}\theta$ factor to be appropriate. In both
cases, the angle is measured from the centre of the sphere. For a conical emitter
rounded at the apex, Spindt et al \cite{Spindt} found the $\theta$ dependence (measured from the
centre of curvature at the tip) to be small close to the
apex though a later study \cite{Li} shows a sharper variation for small $\theta$. 
Clearly, more studies are required to understand the variation of $\gamma$ close to the apex.

The importance of the apex and its immediate neighbourhood arises from the fact that for sharp emitters,
the enhancement factor generally falls rapidly away from the apex even for a decrease in height
by only $R_a$. As a result, the tunneling transmission coefficient can fall by several
orders of magnitude rendering the rest of the emitter inconsequential. The emitter current
can thus be expressed as

\be
I = \int_0^{\rho_{max}} 2\pi \rho \sqrt{1 + (dz/d\rho)^2} J({\bf{r}})~d\rho
\ee

\noi
where ${\bf{r}} = (\rho,z)$ is a point on the emitter surface, $\rho = \sqrt{x^2 + y^2}$ and
$\rho_{max}$ is a cutoff set by accuracy requirements. Here
$J({\bf{r}})$ is the local current density \cite{FN,Nordheim,murphy,jensen2003,forbes_deane,jap2014}
on the emitter surface, calculated by taking into account the local field enhancement factor $\gamma({\bf{r}})$.
The enhancement factor $\gamma({\bf{r}})$ around the apex thus holds the key in any field emission
calculation.

In the following, we shall first study the field enhancement factor for the hemiellipsoid
and cast it in a generalized form $\gamma = \gamma_a \cos{\tilde{\theta}}$ where $\tilde{\theta}$
is defined using normalized co-ordinates ($\tilde{\rho},\tilde{z}$). We shall then deal with
locally quadratic
emitter tips and show numerically that the enhancement factor variation is well described
by this generalization.

\section{Field Enhancement for the Hemiellipsoid}

The vertical hemiellipsoid on a grounded conducting plane placed in an external electrostatic
field ($E_0 \hat{z}$) pointing along the axial direction is one of few analytically solvable models that have
helped in understanding local field enhancement. It is convenient to work in
{\it prolate spheroidal coordinate} system ($\xi,\eta,\phi$). These are related to the
Cartesian coordinates by the following relations:

\bea
\nonumber
&&x= L \sqrt{({\eta^2}-1)(1-{\xi^2})}\cos{\phi}\\
\nonumber
&&y= L \sqrt{({\eta^2}-1)(1-{\xi^2})}\sin{\phi}\\
&&z= L \xi \eta,
\label{Eq:cart_pro_sph}
\eea

\noi
where $L = \sqrt{h^2-b^2}$,  $h$ is the height and $b$ is the radius of the base of the
hemiellipsoid respectively. Note that a surface obtained by fixing $\eta=\eta_0$ in this
coordinate system is an ellipsoid.  For a prolate hemiellipsoid on a grounded plane in the presence of
an external electrostatic field $-E_0 \hat{z}$, the solution of Laplace equation may be written as\cite{kos,jap2016},

\be
V(\eta,\xi)=\xi\eta\left[C'+D'\left({\frac{1}{2}} \ln{\frac{\eta+1}{\eta-1}}-{\frac{1}{\eta}}\right)\right],
\ee

\noindent
where $C'= L E_0$ and  

\be
D'= -L E_0\left({\frac{1}{2}} \ln{\frac{\eta_0+1}{\eta_0-1}}-{\frac{1}{\eta_0}}\right)^{-1}
\ee

\noi
where $\eta=\eta_0$ is the surface of the emitter.

In order to relate this to the enhancement factor, $\gamma$, we need to find the
normal derivative of the potential, $V$ at the surface of the emitter. 
To do so, we first note that

\be
{\bf E}_{local} = - \hat{\eta} \left[{\frac{1}{h_\eta}}{\frac{\partial V}{\partial \eta}}\right]_{\eta=\eta_0}
\ee

\noindent
where 

\be
h_\eta=\sqrt{\frac{L^2}{\eta_0^2-1}(\eta_0^2 - \xi^2)}.
\ee

\noindent
The magnitude of the local electric field normal to the surface $\eta=\eta_0$ is thus given by

\be
E_{local} = -{\frac{\xi}{h_\eta}}\left[C'+{\frac{D'}{2}}\ln{\frac{\eta_0+1}{\eta_0-1}}-{\frac{D'\eta_0}{\eta_0^2-1}}\right]
\ee

\noindent
Note that at the apex of the hemiellipsoid $\xi = 1$. Thus

\be
\frac{\gamma}{\gamma_a}  = \frac{\xi \sqrt{\eta_0^2-1}}{\sqrt{\eta_0^2 - \xi^2}}
\ee

\noindent
Further, with $\xi = z/h$, $L^2 = h^2 - b^2$, $R_a = b^2/h$ and $z^2/h^2 + \rho^2/b^2 = 1$, we have

\be
\gamma = \gamma_a \xi \sqrt{\frac{b^2}{\frac{b^2}{h^2}z^2 + h^2 - z^2}}
\ee

\noi
so that

\be
\gamma = \gamma_a \xi \sqrt{\frac{b^2}{\frac{b^2}{h^2}z^2 + \frac{h^2}{b^2}\rho^2}}
\ee

\noi
and finally

\be
\gamma =  \gamma_a \frac{z/h}{\sqrt{(z/h)^2 + (\rho/R_a)^2}}  \label{eq:variation}.
\ee

\noindent
With $\tilde{z} = z/h$ and $\tilde{\rho} = \rho/R_a$, we define

\be
\cos\tilde{\theta} = \frac{\tilde{z}}{\sqrt{{\tilde{z}}^2 + {\tilde{\rho}}^2}}
\ee

\noi
so that $\gamma = \gamma_a \cos\tilde{\theta}$. In the limit of the hemisphere where
$h = R = R_a$, $\tilde{\theta} = \theta$. Thus, both the hemiellipsoid and hemisphere can be
described by Eq.~\ref{eq:variation}.

\section{Quadratic surfaces}

Generic smooth axially symmetric vertical emitter tips can be described as $z = z(\rho)$.
A Taylor expansion at the apex yields

\bea
z &  = & h  + \frac{1}{2} \Big(\frac{d^2 z}{d\rho^2}\Big)_{\rho = 0} \rho^2  + \ldots \\
& \simeq &  h\Big[1 - \frac{1}{2} \frac{\rho}{R_a}\frac{\rho}{h} \Big] \label{eq:quadratic}
\eea

\noi
where $R_a$ is the magnitude of the apex radius of curvature and $h$ is the height
of the emitter. We have assumed that the
quadratic term is non-zero since $(d^2 z/d\rho^2)_{\rho = 0} = 0$ implies that the tip is
flat rather than having a small radius of curvature characteristic of field emitters.
Also, since field emission occurs close to the tip, we shall ignore higher order terms in $\rho$
as in Eq.~\ref{eq:quadratic}.

The ellipsoid for instance can be expanded as

\be
z = h\Big[1 - \frac{1}{2} \frac{\rho}{R_a}\frac{\rho}{h} - \frac{1}{8} \Big(\frac{\rho}{R_a}\Big)^2\Big(\frac{\rho}{h}\Big)^2 - \frac{1}{16} \Big(\frac{\rho}{R_a}\Big)^3 \Big(\frac{\rho}{h}\Big)^3 - \ldots \Big]
\ee

\noi
which reduces to 

\be
z = R\Big[1  - \frac{1}{2}\Big(\frac{\rho}{R}\Big)^2 - \frac{1}{8}\Big(\frac{\rho}{R}\Big)^4 - \frac{1}{16}\Big(\frac{\rho}{R}\Big)^6 - \ldots\Big]
\ee

\noi
for the sphere. For hemiellipsoidal emitters with large $h$, a quadratic truncation seems adequate.

Such quadratic emitter tips can thus be considered generic for purposes of field emission.
They may be mounted on a variety of bases, ranging from the classical cylindrical
post typical of carbon nanotubes to the conical bases of a Spindt array \cite{Spindt}
or even be part of compound structures.
We shall study the applicability of Eq.~\ref{eq:variation} for such emitter tips numerically.

\section{Numerical Studies}

We shall adopt the {\it{nonlinear}} line charge model \cite{jap2016,harris_inf3} to determine the electrostatic potential
and thus the field enhancement factor. It consists of a vertically placed line charge of height ${\cal L}$ on
a grounded plane in the presence of an external electrostatic field $E_0$. The line charge together with
its image and the external field produces a zero-potential surface that coincides with the emitter surface under
study. The shape of the zero-potential surface crucially depends on the line charge density. Thus for
a linear line charge density the shapes generated are hemiellipsoidal, while non-linear line charge densities
can generate a wide variety of shapes including a conical base with a quadratic top.

For our purposes, we consider a polynomial line charge density $\Lambda(z) = \sum_{n=0}^{N} c_n z^n$
with the coefficients $c_n$ chosen appropriately. The potential at a point external to the
emitter can thus be expressed as

\bea
V(x,y,z) & = &  \frac{1}{4\pi\epsilon_0} \Big [ \int_{0}^{L} ds \frac 
{\Lambda(s)}{(\rho^2 + (z-s)^2)^{1/2}} \\ \nonumber
& - & \int_{0}^{L} ds \frac{\Lambda(s)}{(\rho^2 + (z+s)^2)^{1/2}} \Big ] -  E_0 z
\eea

\noi
where $\rho^2 = x^2 + y^2$. The local electrostatic field, ${\bf{E}} = -\nabla(V)$, can thus be used
to determine the field enhancement factor $\gamma$ on the surface of the emitter.
We have validated the method successfully using the hemiellipsoidal emitter over a range of aspect
ratios. We have chosen a variety of emitter shapes for our study with parameters values that allow a
span of the apex enhancement factor from a few 100s to 12000.  

\begin{figure}[hbt]
  \begin{center}
    \vskip -0.5cm
\hspace*{-1.0cm}\includegraphics[scale=0.36,angle=0]{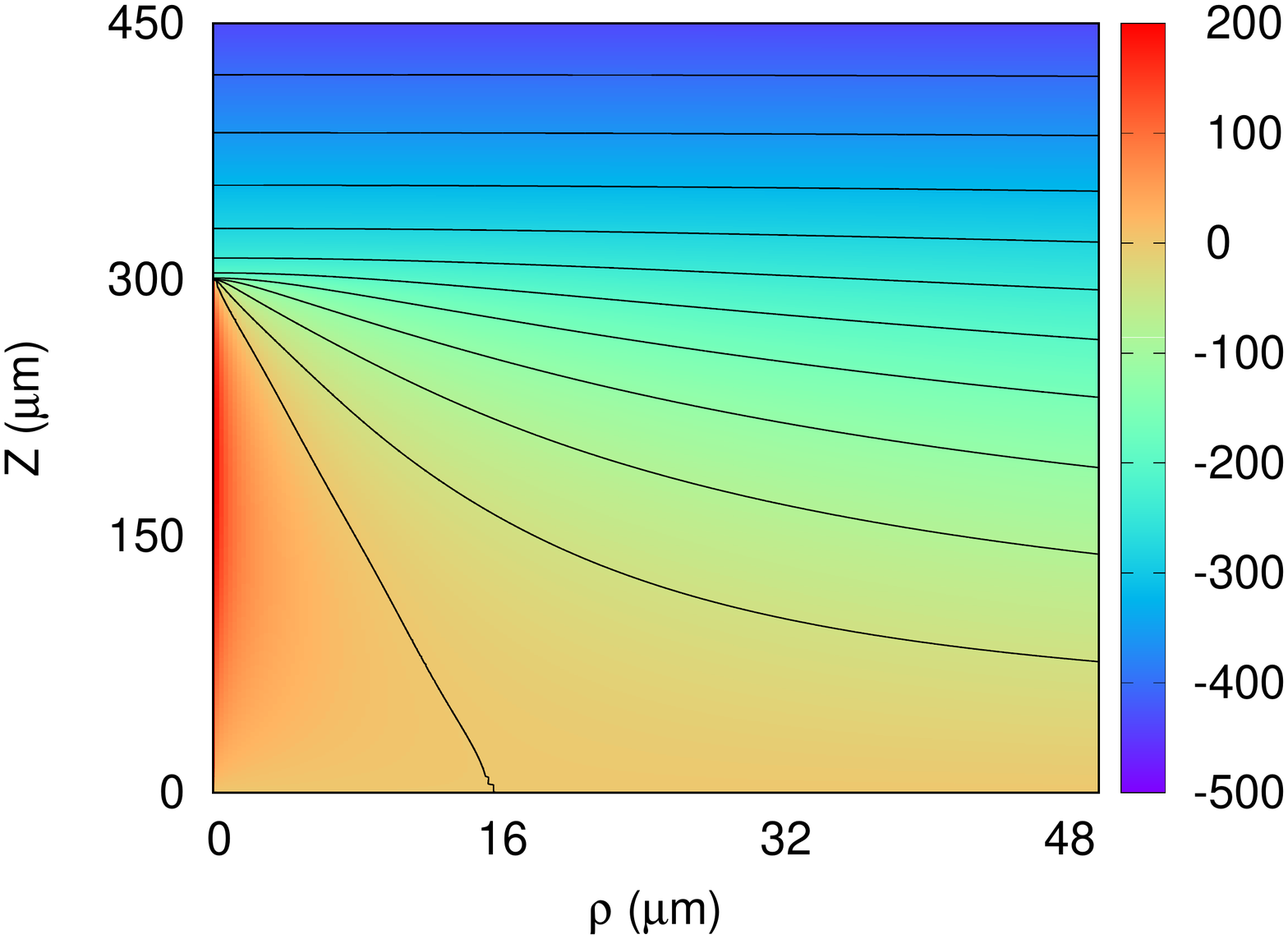}
\caption{A contour plot of the potential for a grounded conical emitter of height $h = 300~\mu$m and
  base radius $b = 16~\mu$m in an external electrostatic field $E_0 = 1$~MV/m. }
\label{fig:cone_contour}
\end{center}
\end{figure}

\begin{figure}[h]
  \begin{center}
    \vskip -0.5cm
\hspace*{-1.0cm}\includegraphics[scale=0.36,angle=0]{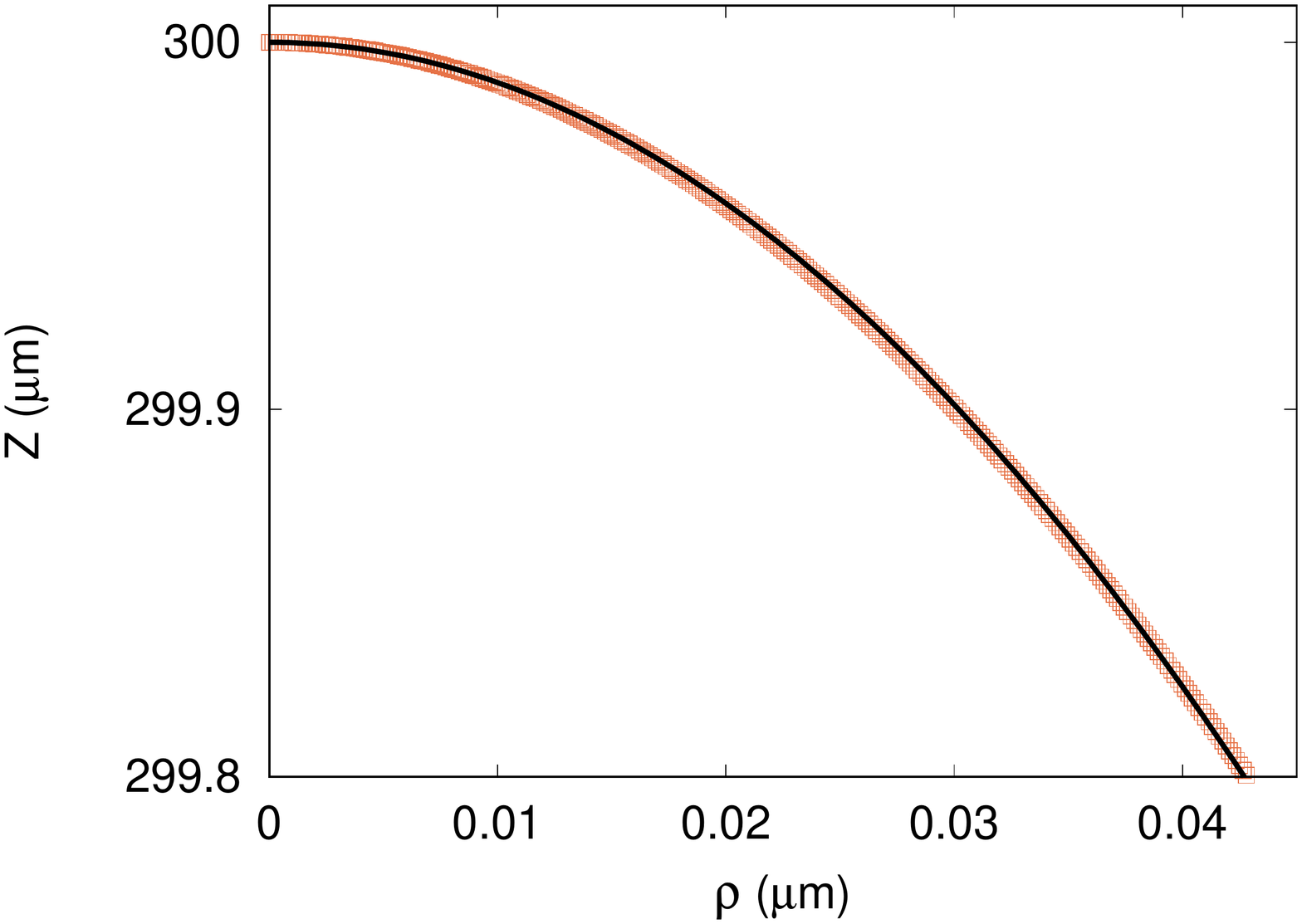}
\caption{The conical tip is rounded due to the nature of the line charge distribution. The
apex radius of curvature $R_a = 4.56$~nm in this case. A quadratic $z = h - a \rho^2$ (solid line) fits well
  in the neighbourhood of the tip.}
\label{fig:cone_apex}
\end{center}
\end{figure}

\begin{figure}[hbt]
  \begin{center}
    \vskip -0.5cm
\hspace*{-1.0cm}\includegraphics[scale=0.36,angle=0]{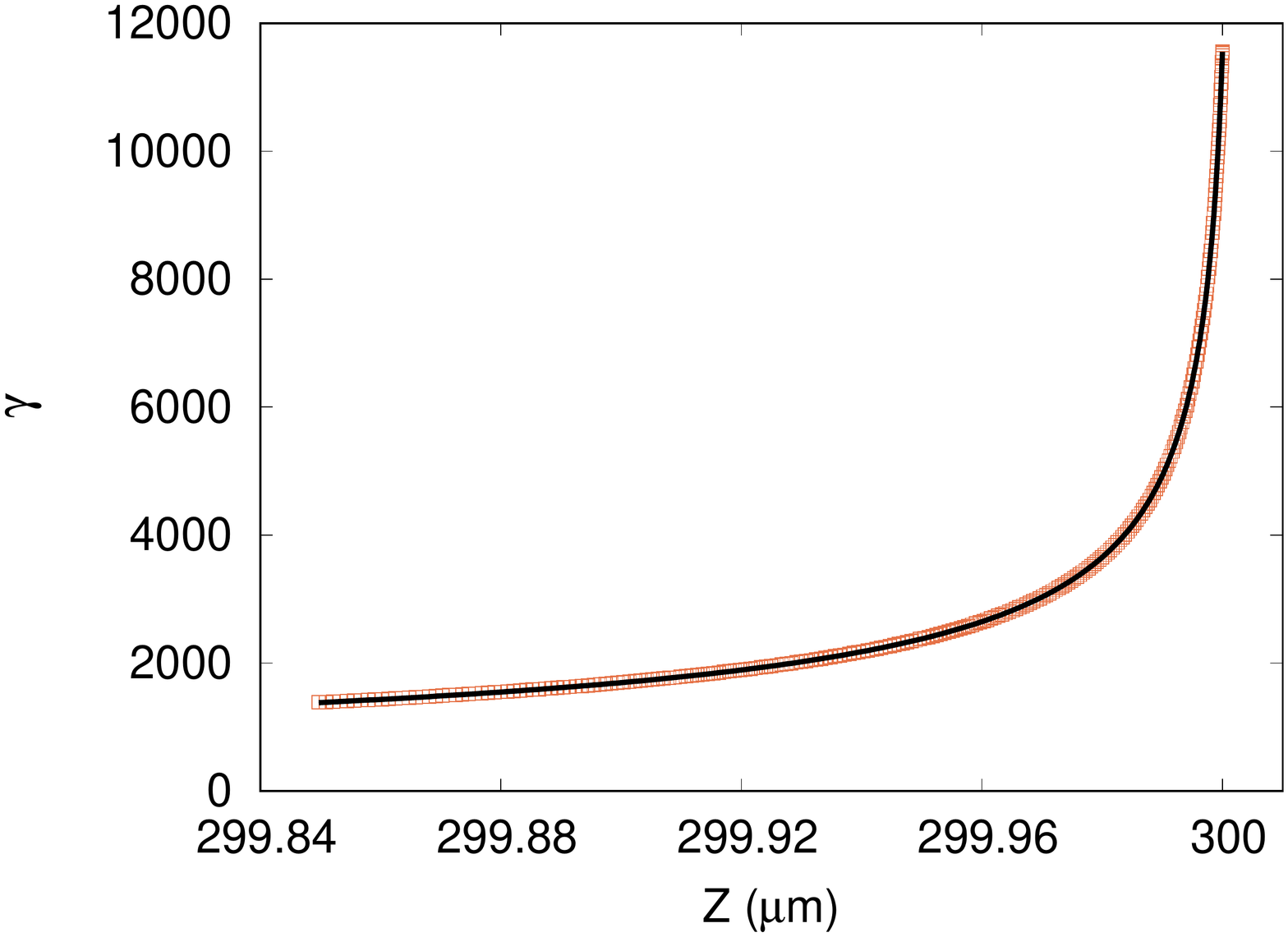}
\caption{The variation of the enhancement factor $\gamma$ near the tip of the cone
  is compared with $\gamma_a \cos\tilde{\theta}$.}
\label{fig:vary_cone}
\end{center}
\end{figure}

Fig.~\ref{fig:cone_contour} shows the potential profile of a single circular cone-shaped emitter on an infinite 
grounded metallic plate, placed in an external field ${\bf{E}} = \hat{z} E_0$ where $E_0 = 1$~MV/m.
The height of the cone is $300~\mu$m while the base radius $b = 16~\mu$m. 
The line charge density is such that the tip is rounded at the apex. The zero contour profile
near the tip is shown in Fig.~\ref{fig:cone_apex} along with a quadratic fit $z = h - a_2 \rho^2$
with $a_2 = 109.65~\mu{\rm m}^{-1}$. The excellent agreement shows that the region near the tip
is locally a quadratic surface, with the apex radius of curvature $R_a = 1/(2a_2) \simeq 4.56$~nm.
The variation of the enhancement factor $\gamma(z)$ near the apex is shown in Fig.~\ref{fig:vary_cone}.
along with the curve $\gamma_a \cos\tilde{\theta}$ (see Eq.~\ref{eq:variation}), calculated using
the height $h$ and the apex radius of curvature $R_a$. The agreement shows that, close to the
apex, the variation in the field enhancement factor is well described by the formula derived
for an hemiellipsoid.

\begin{figure}[hbt]
  \begin{center}
    \vskip -1.75cm
\hspace*{-1.0cm}\includegraphics[scale=0.36,angle=0]{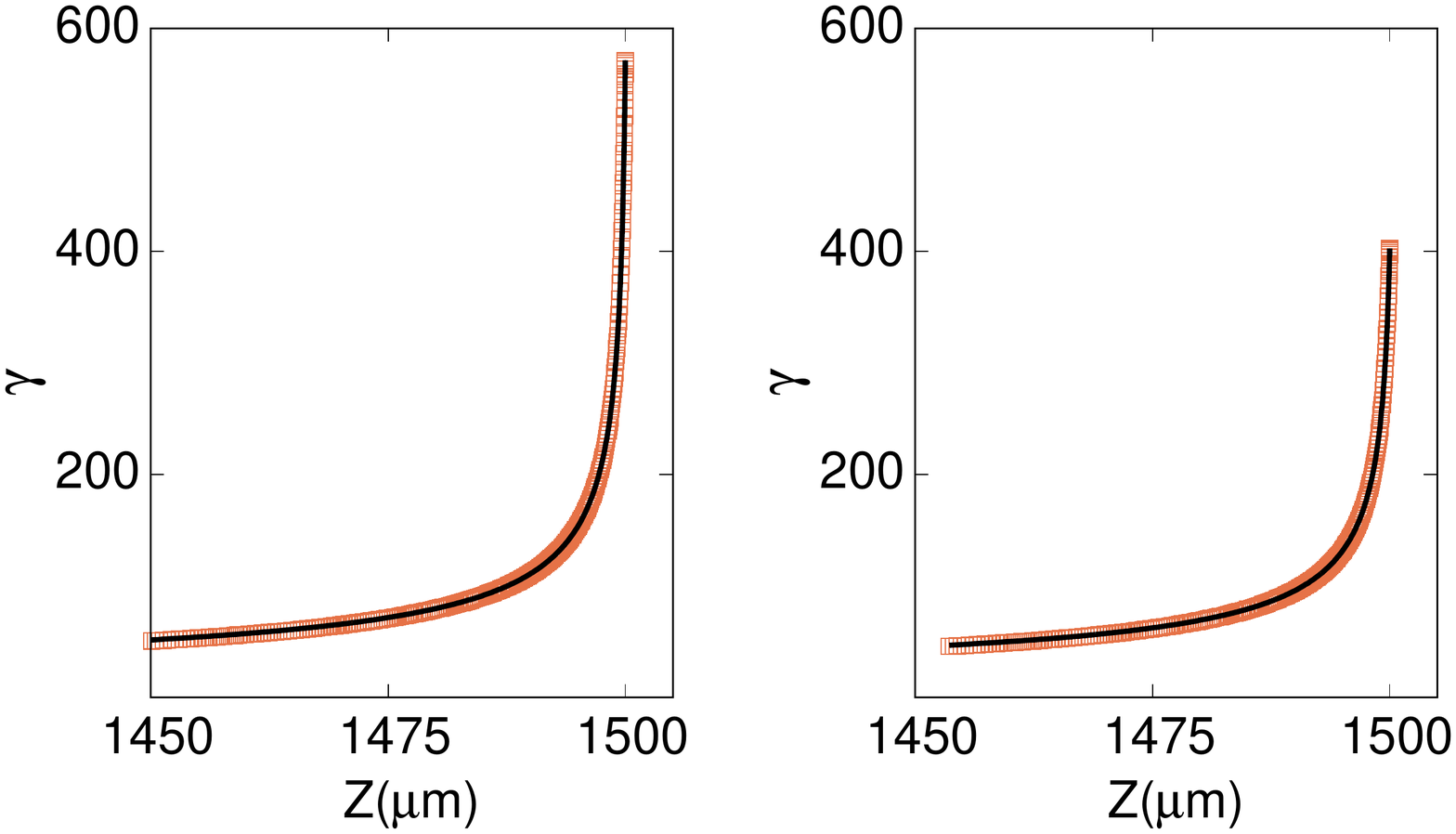}
\caption{ The variation of the enhancement factor $\gamma$ near the tip of 2 surfaces of
  height 1500~$\mu$m and base radius 20~$\mu$m, having apex radius of curvature
  0.77~$\mu$m (left) and 1.18~$\mu$m (right). Their apex enhancement factors (the value of $\gamma$ at $z=1500~\mu$m)
  are 571 and 402 respectively. For both cases, Eq.~\ref{eq:variation} is also plotted  (solid lines) and found to 
describe the variation well.}
\label{fig:vary_P3n5}
\end{center}
\end{figure}

We next consider 2 surfaces of height $1500~\mu$m and base radius $b = 20~\mu$m with a quadratic tops.
In the first case, the
apex radius of curvature $R_a = 0.77~\mu$m while for the second $R_a = 1.18~\mu$m.
The apex enhancement factor are $\gamma_a = 571$ and
$\gamma_a = 402$ respectively. The variation of the enhancement factor $\gamma$ is shown in
Fig.~\ref{fig:vary_P3n5}. In both cases, Eq.~\ref{eq:variation} describes the variation well.

\begin{figure}[hbt]
  \begin{center}
    \vskip -1.25cm
\hspace*{-1.0cm}\includegraphics[scale=0.36,angle=0]{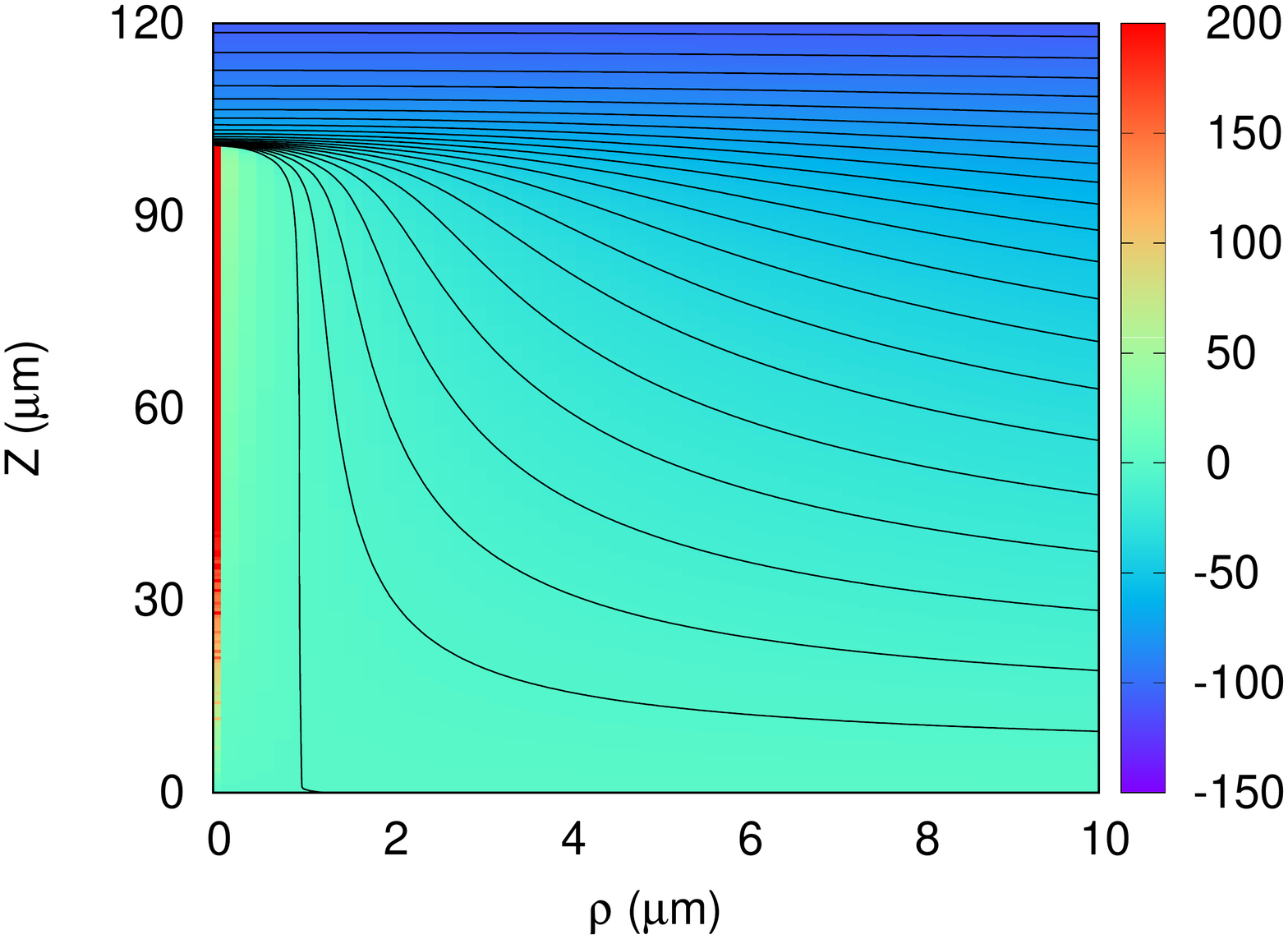}
\caption{A contour plot of the potential for a grounded emitter shaped as a cylindrical post with
  a generic top. The total height $h = 101~\mu$m while the
  base radius $b = 1~\mu$m and the external electrostatic field $E_0 = 1$~MV/m. The apex
enhancement factor $\gamma_a = 285$.}
\label{fig:hemi_post}
\end{center}
\end{figure}

\begin{figure}[hbt]
  \begin{center}
    \vskip -0.5cm
\hspace*{-1.0cm}\includegraphics[scale=0.36,angle=0]{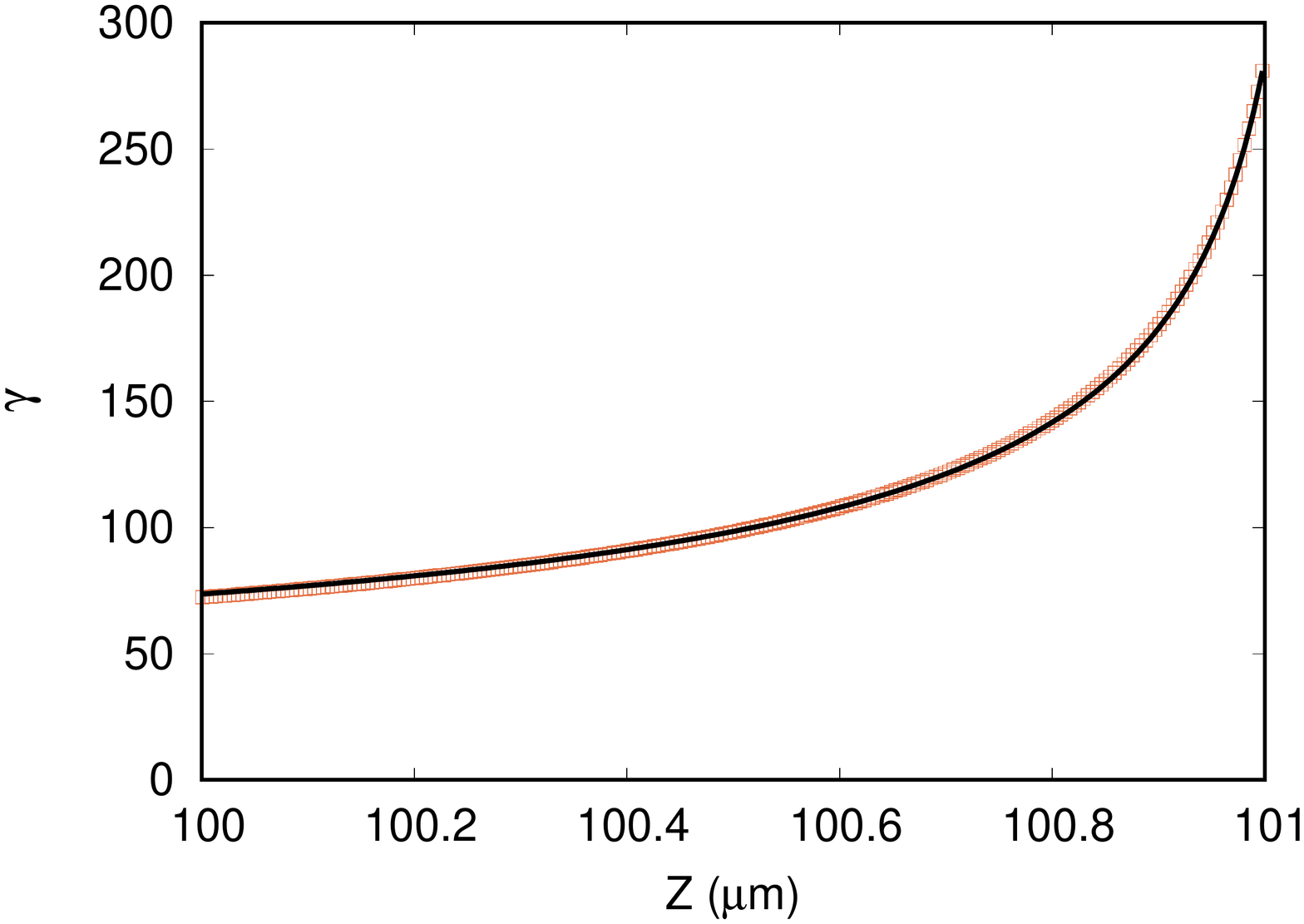}
\caption{ The variation of the enhancement factor $\gamma$ near the tip for the cylindrical
  post with a locally quadratic non-spherical top. The apex enhancement factor is 285.
  The solid line is Eq.~\ref{eq:variation} with $h = 101~\mu$m and $R_a = 0.1278~\mu$m.
}
\label{fig:vary_hemi}
\end{center}
\end{figure}

As a final example, we consider a cylindrical post with a top that is locally quadratic, typical
of carbon nanotubes. The height of the system is $101~\mu$m while the
base radius is $1~\mu$m. The potential profile in
the presence of an external electrostatic field $E_0 = 1$~MV/m, is shown in Fig.~\ref{fig:hemi_post}.
The apex radius of curvature $R_a = 0.1278~\mu$m. Fig.~\ref{fig:vary_hemi} shows the variation
of the enhancement factor $\gamma$. As before, Eq.~\ref{eq:variation} describes the variation
in enhancement factor very well.

The variation in field enhancement factor expressed in Eq.~\ref{eq:variation} is thus
found to hold for a variety of shapes apart from the hemiellipsoid for which it was derived.
We have tested its veracity for numerous other shapes that have not been presented here,
including compound shapes formed by mounting an emitter on a pedestral such as a cylinder.

\section{Discussion and Conclusion}

Before summarizing our results, we make a couple of  observations. 
First, we believe the results presented here to be generic and hence applicable to a wide
class of emitters provided the anode is flat and the anode-cathode gap large compared to the height
of the emitter. There is however a discrepancy with the numerical results reported
for the hemisphere on a cylindrical post \cite{podenok,read,roveri} where the surface variation of $\gamma$
seems to differ from Eq.~\ref{eq:variation} and appears to be qualitatively closer to the numerical
result presented in Fig.~4 of [13] for a hemiellipsoid \cite{ellip,notdone}.
It may also be noted that result of Dyke et al \cite{dyke53} for a cone with a spherical top
does not fall within the ambit of the present work since the anode in their work \cite{dyke53}
belongs to the same family as the cathode.

Second, a somewhat related problem is the distribution of excess charge on the surface of a
conductor and its relation to the local curvature of the conductor. In the {\it absence of
an external field}, the surface charge density $\sigma \sim \kappa_G^{1/4}$ for quadratic surfaces
(such as an ellipsoid, paraboloid or a hyperboloid) where $\kappa_G$ is the local Gaussian
curvature \cite{gaussian}. For an ellipsoid,

\be
\kappa_G = \frac{1}{R_a^2} \frac{1}{ \big[(z/h)^2 + (\rho/R_a)^2\big]^2}
\ee

\noi
so that

\be
\sigma \sim \frac{1}{ \sqrt{(z/h)^2 + (\rho/R_a)^2}}.
\ee

\noi
For $z \sim h$, this is similar to Eq.~\ref{eq:variation} which describes the local field variation on the
surface of a conductor, though in the presence of an external field ${\bf{E}} = -\hat{z} E_0$.

In conclusion, we have studied the variation in the field enhancement factor on the surface
of a conductor close to it apex, in the presence of an external electrostatic field along the
symmetry axis ($\hat{z}$). For the hemiellipsoid, we have expressed the variation exactly over the entire
surface as a generalized
$\cos\tilde{\theta}$ factor, similar to the hemisphere on a conducting plane. We have numerically
tested the validity of the $\cos\tilde{\theta}$ factor extensively for other emitter shapes and found it to
describe the variation effectively near the apex. Since the enhancement factor falls sharply away
from the apex, such a description is adequate for purposes of field emission. Thus, a simplified
description of emitters consists of tips that can be expressed as $z = h  -  \frac{1}{2R_a}\rho^2$,
where both the height $h$ and apex radius $R_a$ are  experimentally measurable parameters which can in turn 
be used to predict the field enhancement variation.

\vskip 0.05 in
$\;$\\
\section{References} 



\begin{thebibliography}{99}
\bibitem{choice} The cylindrical post with a hemispherical top has been the model of choice for studying
nanotubes \cite{bonard,smith} though differently shaped  emitter tips mounted on a
cylinderical post are perhaps more appropriate.
\bibitem{ellip} A hemiellipsoid should follow Eq.~\ref{eq:variation} exactly provided the anode is
  flat and sufficiently far away from the cathode. We have verified this
  numerically as well using the line charge model. The results in [13] (in particular Fig.~4) for the
  ellipsoid therefore seems to be an aberration.
\bibitem{notdone} We have not been able generate to suitable nonlinear line charge density to
  model a cylindrical post with an exact hemispherical top.
\end{thebibliography}
\end{document}